\documentclass[preprint,amsmath,amssymb,superscriptaddress,floatfix,footinbib]{revtex4-2}

\usepackage[english]{babel}
\usepackage{graphicx} 
\usepackage{bm}
\usepackage{float}
\usepackage{bbm}
\usepackage{color}
\usepackage{amsmath}
\usepackage{amssymb} 
\usepackage{color}
\usepackage{footnote}
\usepackage{comment}
\usepackage{hyperref}
\usepackage{subfigure}
\usepackage{appendix}
\usepackage[applemac]{inputenc}
\usepackage[T1]{fontenc}
\usepackage{physics}
\usepackage[english]{babel}  
\usepackage{caption}
\usepackage{ragged2e} 

\usepackage{caption}
\newcommand{\Ca}{Ca$_{2}$RuO$_{4}$}
\newcommand{\CaMn}{Ca$_{2}$Ru$_{1-x}$Mn$_x$O$_{4}$}

\begin{document}


\title{Unveiling Hidden Magnons with Anomalous Rotational Symmetry}


\author{Dirk Wulferding$^*$}
\affiliation{Department of Physics and Astronomy, Sejong University, Seoul 05006, Republic of Korea}

\author{Francesco Gabriele$^*$}
\affiliation{CNR-SPIN, c/o Universit\`a di Salerno, IT-84084 Fisciano (SA), Italy}

\author{Wojciech Brzezicki}
\affiliation{Institute of Theoretical Physics, Jagiellonian University, ulica S. \L{}ojasiewicza 11, PL-30348 Krak\'ow, Poland} 
\affiliation{International Research Centre MagTop, Institute of Physics, Polish Academy of Sciences, Aleja Lotnik\'ow 32/46, PL}

\author{Mario Cuoco}
\affiliation{CNR-SPIN, c/o Universit\`a di Salerno, IT-84084 Fisciano (SA), Italy}

\author{Changyoung Kim}
\affiliation{Center for Correlated Electron Systems, Institute for Basic Science, Seoul 08826, Republic of Korea}
\affiliation{Department of Physics and Astronomy, Seoul National University, Seoul 08826, Republic of Korea}

\author{Mariateresa Lettieri}
\affiliation{CNR-SPIN, c/o Universit\`a di Salerno, IT-84084 Fisciano (SA), Italy}

\author{Anita Guarino}
\affiliation{CNR-SPIN, c/o Universit\`a di Salerno, IT-84084 Fisciano (SA), Italy}

\author{Antonio Vecchione}
\affiliation{CNR-SPIN, c/o Universit\`a di Salerno, IT-84084 Fisciano (SA), Italy}

\author{Rosalba Fittipaldi}
\affiliation{CNR-SPIN, c/o Universit\`a di Salerno, IT-84084 Fisciano (SA), Italy}

\author{Filomena Forte}
\affiliation{CNR-SPIN, c/o Universit\`a di Salerno, IT-84084 Fisciano (SA), Italy}

\begin{abstract}
Correlated materials with competing spin--orbit and crystal-field interactions can host composite spin--orbital magnons that are highly susceptible to structural and electronic perturbations, enabling the control of magnetic dynamics beyond spin-only physics. Using Raman spectroscopy on Ca$_2$RuO$_4$, we show that the partial substitution of Ru with Mn reconstructs the magnon spectrum and reveals one-magnon modes that are hidden in the undoped state. We demonstrate that the transition-metal substitution activates otherwise 
symmetry-forbidden magnon modes 
through mirror-symmetry breaking of the underlying spin--orbital configuration. This effect can be theoretically explained by the local structural distortions induced in the RuO$_6$ octahedra near the dopant, that enable the observation of mixed-parity one-magnon modes. These excitations display a distinctive polarization dependence, with a lowering from fourfold to twofold rotational symmetry arising from the mixed-parity character of the coupled magnons and interference between resonant and nonresonant scattering channels. Our results show that spin--orbit--lattice entanglement provides a route to tailoring collective magnetic excitations and their polarization response in spin--orbit--coupled correlated systems.
\end{abstract}
\maketitle

\noindent  

\section{Main}

Spin--orbit coupling (SOC) in correlated electron systems can fundamentally reshape magnetic ground states and excitations, giving rise to phenomena that go beyond conventional spin-only descriptions. When SOC, crystal-field splittings, and electron correlations are of comparable magnitude, spin, orbital, and lattice degrees of freedom become strongly entangled, enabling unconventional forms of magnetism and collective excitations~\cite{Georges2013,Khaliullin2005,Cao2018,Takayama2021,Pesin2010}. Understanding how these intertwined interactions respond to controlled perturbations is a central challenge in the study of quantum materials.

Ca$_2$RuO$_4$ is a prototypical spin--orbit--coupled Mott insulator that exhibits anomalous ground-state responses and a rich phase diagram \cite{Alexander1999PRB,Nakatsuji2000doping, Ricco2018strain, Friedt2001,Cuono2022,Gauquelin2023NanoLett,Cuono_2025} arising from the interplay of spin, orbital, and lattice degrees of freedom \cite{SutterNatComm2017a,das_spin-orbital_2018, Nakamura2023UniqueItinerant,jain2017higgs,Gauquelin2023NanoLett,Brzezicki2023}, with still debated properties particularly about the metal--insulator transition under nonequilibrium conditions \cite{Nakamura2013Electricfield, IchiroJSPS2020, Bertinshaw2020UniqueState, Cirillo2019EmergenceCa2RuO4,Curcio2023Current-drivenCa2RuO4,Suen2023NatureTransport-ARPES, Zhang_PRX_2019,Fursich2019RamanCa2RuO4,bhartiya2025evidenceelectronicstatesdriving}.
Below the metal--insulator transition temperature, the distortions of the RuO$_6$ octahedra become more pronounced, and an antiferromagnetic insulating phase emerges with a N\'{e}el temperature $T_N \approx 110$~K~\cite{Nakatsuji1997,Cao1997,Braden1998,Friedt2001}. While initially described in terms of localized $S=1$ moments of Ru$^{4+}$ ($4d^4$), it is now understood that the magnetism of Ca$_2$RuO$_4$ arises from a competition between SOC and crystal-field effects, leading to a spin--orbital--entangled ground state~\cite{jain2017higgs,das_spin-orbital_2018,souliou_raman_207,porterPRB2018,Fatuzzo2015,GretarssonPRB2019}. Ca$_2$RuO$_4$ is thus close in proximity to a $d^4$ excitonic magnet, in which long-range order emerges from the coupling of spin--orbit excitons rather than from preformed local moments~\cite{Khaliullin2013,fortePRB2010,Feldmaier2020,Sarte2020}. Magnetic excitations in Ca$_2$RuO$_4$ reflect this entanglement, exhibiting strong anisotropy, sizable excitation gaps, and modes with mixed spin--orbital character, as revealed by neutron scattering, Raman and resonant inelastic scattering spectroscopies~\cite{jain2017higgs,das_spin-orbital_2018}. A challenging feature of this material is the sensitivity of its magnetic properties to lattice distortions. Changes induced by temperature, pressure, strain, or chemical substitution can reconstruct the magnetic ground state~\cite{Steffens2005,Qi2010,porterPRB2018,Chi2020,Porter2022,Tian2024,Kunkemoller2017,Pincini2018,Brzezicki2015,Brzezicki2020}. This pronounced spin--orbital--lattice coupling makes Ca$_2$RuO$_4$ an ideal platform for exploring how symmetry breaking and structural perturbations can be exploited to tune collective magnetic dynamics in spin--orbit--coupled correlated materials.

The observability of collective magnetic excitations is governed by crystal symmetry and experimental selection rules. As a consequence, a portion of the magnon spectrum may remain hidden, being symmetry--forbidden.
Optical and neutron--scattering studies have revealed that structural perturbations, reduced dimensionality, or hybridization with lattice degrees of freedom can render otherwise forbidden one--magnon modes observable \cite{Li1987,Etesamirad2023}. 
Hybrid magnon--phonon excitations \cite{Metzger2024} and nonreciprocal magnons \cite{Gitgeatpong2017} further illustrate how subtle symmetry breaking \cite{Winter2017,Kim2022,Chen2025} can dramatically reshape magnetic excitation spectra. To probe these spectral changes, polarization--resolved Raman spectroscopy serves as a powerful tool, as the polarization of light encodes the symmetry of collective excitations \cite{Zhang2019-th, McCreary2020,Glamazda2016,Yang2021}. Particularly under resonant conditions, additional scattering pathways can substantially reshape the Raman response and its symmetry signatures, providing direct insight into modifications of both the magnetic and light--matter interactions.\\ Despite these insights, controlled pathways to activate hidden magnons in bulk spin--orbit correlated materials, without inducing a global structural phase transition remain scarce. Addressing this gap is crucial to establishing spin--orbital--lattice engineering as a viable route for controlling collective magnetic excitations in correlated materials. In this manuscript, we exploit the design path  
of transition-metal substitution in Ca$_2$RuO$_4$ and demonstrate that it can reconfigure the spin--orbital magnon spectrum, leading to the emergence of 
one-magnon modes hidden in the pristine material. Our Raman results, supported by theoretical modeling, reveal that these hidden one-magnon excitations are activated by the breaking of mirror symmetry within the spin--orbital manifold. This effect, driven by local RuO$_6$ octahedral distortions near the dopants, induces the mixing of even- and odd-parity states, thereby enabling the realization of mixed-parity magnon modes. We further uncover an unconventional polarization dependence of the magnon modes emerging upon transition-metal substitution, characterized by a reduction from fourfold to twofold rotational symmetry. We attribute this lowered symmetry to the mixed-parity character of the coupled magnons and to interference between resonant and nonresonant scattering channels.  


\section{Evidence of hidden magnons and rotational symmetry breaking}

To investigate the impact of Mn substitution on the magnetic excitations of Ca$_2$RuO$_4$, we perform Raman spectroscopic measurements at cryogenic temperatures. Raman spectroscopy is a useful technique for directly probing magnetic excitations in ruthenates in the form of one- and multi-magnon modes \cite{souliou_raman_207,lee23,wulferding-23}, whose energies are directly related to the magnetic correlations between Ru ions.
Ca$_2$RuO$_4$ belongs to the orthorhombic space group Pbca (\#61). For Raman measurements within the crystallographic $ab$ plane, this space group yields excitations of $A_g$ and of $B_{1g}$ symmetry. 
In Fig. 1a we plot temperature dependent Raman data taken on Ca$_2$(Ru,Mn)O$_4$ with 0\%, 3\%, 5\%, and 10\% Mn content. The outcomes at 0\% are completely consistent with those reported in literature \cite{souliou_raman_207}. We can clearly identify $T_N$ for each compound: at high temperatures ($T>T_N$) there is an enhanced scattering background due to paramagnetic fluctuations. These fluctuations are quenched below $T_N$, resulting in a gapped background (this is most clearly seen for the sample with 0\% Mn below $\sim 60$ meV). Simultaneously, new one-magnon peaks, marked by arrows in each panel, emerge below $T_N$. For a better comparison of the evolution of phononic and magnetic excitations with increasing Mn content, we plot individual spectra of all four compounds obtained at a base temperature of $T = 4$ K in Fig. 1b. One-magnon modes are shaded, while partially overlapping phonons are marked by asterisks. The pure Ca$_2$RuO$_4$ sample hosts a sharp and intense single one-magnon excitation at 12.5 meV (yellow-shaded), which is consistent with the previous report~\cite{souliou_raman_207}. A partial substitution of Ru with 3\% and 5\% Mn yields two one-magnon branches with reduced intensity and increased linewidth (shaded green and red), while for a larger Mn content of 10\% only a single, broadened and asymmetric magnon mode (yellow-shaded) can be resolved. The extracted parameters and energies are summarized in Table I in the Supplementary Information, and one-magnon energy and intensity as a function of Mn content are plotted in Figs. 1c and 1d.
Consequently, even a small level of substitution shifts the one-magnon mode to higher energies and leads to the emergence of an additional magnetic peak, with a maximum intensity at an energy separation of approximately 1 meV. Further increases in Mn concentration continue to raise the magnon energy and result in a pronounced broadening of the spectral intensity, such that the two split peaks can no longer be distinguished.
We further note that Mn--Ru substitution gives rise to additional low--energy magnetic excitations characterized by weak intensity and a broad continuum extending from approximately 5 to 8 meV. The associated spectral weight shows a slight increase with increasing Mn concentration.
In addition to the low-energy magnetic modes, an extra phononic peak emerges near the phonon excitations at around 31 meV and 48 meV as a consequence of Mn substitution. By contrast, no significant modifications are observed in the higher-energy features within the 70--90 meV range.\\
Here, we concentrate on the one-magnon excitations in the cross-polarization channel, as they exhibit the most pronounced reconstruction upon Ru substitution.

A deeper insight into the symmetry properties of all observed Raman-active excitations can be gained from polarization-resolved experiments. Fig. 2a illustrates three fundamental scattering configurations by relating the polarization vectors of incident and outgoing light to the crystal orientation. In the color contour plot shown in Fig. 2b we detail the evolution of the Raman scattering intensities for individual excitations as a function of light polarization, measured on the 3\% Mn sample in crossed polarization (i.e., $B_{1g}$-selective). Note that all phonon modes (17 meV and higher) show a four-fold periodicity, as expected for modes of $B_{1g}$ symmetry.

In contrast, the two lowest-lying excitations, corresponding to the two branches of one-magnon excitations and marked by green and red arrows, deviate from this behavior and display a two-fold distortion on top of their four-fold periodicity. To emphasize this distortion, and to compare it among the four different samples of our study, we extract the intensity linecuts as shown in the polar plots of Figs. 2c-f. Strikingly, in samples where both magnon branches are resolved (i.e., in the 3\% and 5\% Mn samples), we find a distinct alternating, out-of-phase behavior for the intensity profiles of the two magnons, which may be taken as a consequence of the mixed-parity character of the magnons.

Magnetic excitations of $B_{1g}$ symmetry adhere to a four-fold, $d$-wave-like intensity profile, where the scattering process is fully described by the Loudon-Fleury (LF) theory \cite{Fleury1968}. Conversely, in the presence of additional, non-Loudon-Fleury terms \cite{shastry1990,Devereaux2007,Chubukov1995,Yang2021}, this symmetry may be lowered towards a two-fold distorted pattern. Such non-Loudon-Fleury terms may be significantly enhanced through resonance processes. 
To uncover the presence and relevance of these non-Loudon-Fleury contributions, we therefore compare the observed one-magnon intensity profiles probed with a $\lambda=561$ nm laser ($E_{\mathrm{photon}}=2.21$ eV) to profiles measured with a $\lambda=660$ nm laser ($E_{\mathrm{photon}}=1.87$ eV). We quantify the degree of deviation from conventional LF theory by taking the ratio of lobe intensities measured along the 0$^{\circ}$-to-180$^{\circ}$ axis versus intensities measured along the 90$^{\circ}$-to-270$^{\circ}$ axis. These two configurations are defined as $B_{1g}^{(1)}$ and $B_{1g}^{(2)}$, corresponding to the cases in which the incident light polarization vector is rotated by 90$^\circ$ with respect to the crystal axes (Fig.~2a). The resultant ratios are plotted in panels 2g-j for samples with 0\%, 3\%, 5\%, and 10\% Mn, respectively, measured at $T = 4$ K. Here, a ratio of 1 corresponds to a symmetrical four-fold pattern, while a ratio different from 1 signifies the presence of non-LF terms. Remarkably, we find that all samples, as well as both one-magnon branches (where resolved) display their own distinct resonance behavior.


\section{Modelling magnon spectra}

In \Ca\  the Ru$^{4+}$ ions host four electrons in the $t_{2g}$ shell, giving rise to local orbital and spin moments $L=1$ and $S=1$. At the atomic level, the intraionic SOC term $\lambda\,\mathbf{L} \cdot \mathbf{S}$ stabilizes a nonmagnetic $J=0$ ground state, separated from the excited $J=1$ and $J=2$ multiplets. The competition between SOC and electron--lattice interactions --- notably the flattening, rotation, and tilting of the RuO$_6$ octahedra --- drives the system away from the ideal $J=0$ configuration. Consequently, the low-energy manifold forms a quasi-triplet composed of the $J=0$ singlet and the two lowest components of the $J=1$ triplet \cite{Khaliullin2013}. These states define an effective local moment $\mathbf{T}$ with total pseudospin $T=1$, schematically illustrated in Fig.~\ref{fig:ConceptFig}a in terms of its spin--orbital composition.

Within this pseudospin basis, the nearest-neighbor exchange interactions in the RuO$_2$ planes can be constructed by symmetry. For bonds along the $x$ direction, the effective Hamiltonian reads
\begin{align}
\mathcal{H}_x &=
\sum_{\langle i,j\rangle \parallel x}
\Big[
  J_{xx}\, T_i^x T_j^x
  + J_{xy}\, T_i^y T_j^y
  + J_z\, T_i^z T_j^z
\Big],
\label{eq:exchange}
\end{align}
where $J_{xx}$, $J_{xy}$, and $J_z$ are the anisotropic exchange couplings allowed by the orthorhombic symmetry. 
An analogous form $\mathcal{H}_y$ applies for bonds along $y$, with $J_{yy}=J_{xx}$ and $J_{yx}=J_{xy}$.

Local symmetry-lowering distortions, including the flattening, rotation and tilting of the RuO$_6$ octahedra, introduce additional on-site staggered contributions (Fig. 2b) \cite{Liu2019}

\begin{align}
\mathcal{H}_{\mathrm{loc}} &=
E_T \sum_j (T_j^z)^2
+ \delta_c \sum_j (-1)^j
  ( T_j^x  T_j^y +T_j^y  T_j^x )
\nonumber\\
&\quad
+  \delta_b \sum_j (-1)^j 
 [ T_j^z (T_j^x+ T_j^y )+ (T_j^x+ T_j^y ) T_j^z ],
\label{eq:Hloc}
\end{align}
where $E_T$ is the single-ion anisotropy due to axial compression along the $c$ axis, and $\delta_c$, $\delta_b$ represent the anisotropy terms associated with staggered octahedral rotations and tilts about the $c$ and $b$ axes, respectively. X-ray 
studies on Mn-doped \Ca\ report local structural relaxations around the impurity sites, which in turn lead to concomitant modifications of the magnetic order \cite{Porter2022}. In our model, the effects of impurities at dilute concentrations mainly affect the local crystal field environment. Specifically, since Mn enters with a $d^3$ non-Jahn Teller configuration, the substitution tends to symmetrize the octahedral Ru-O bonds by reducing the tetragonal energy splitting $\Delta_T$. Hence, we effectively assume that the Mn dopant amplifies the amplitude of the $E_T$ term, as it scales as $\lambda^2/\Delta_T$ \cite{das_spin-orbital_2018}. 

We now describe the symmetry aspects of the low energy magnon modes that are crucial to account for their reconstruction upon Mn substitution. The low-energy spin-orbital states of the $t_{2g}$ manifold ($L=1, S=1$) couple anisotropically to the lattice via octahedral compression, rotation about the $c$ axis, and tilting relative to the $b$ axis. These states can be classified by their parity under the vertical mirror operation $\mathcal{M}_b$
(Fig.~\ref{fig:ConceptFig}a). Specifically, the ground-state singlet (adiabatically connected to the $J=0$ state) is $\mathcal{M}_b$-odd, whereas the excited doublet ($J=1$ subspace) comprises states of opposite parity. In the effective magnetic exchange model, these map to the $T=1$ states $\lbrace\ket{\tilde{1}}, \ket{\tilde{0}}, \ket{-\tilde{1}}\rbrace$, which diagonalize the local Hamiltonian of Eq. \ref{eq:Hloc} and satisfy the aforementioned symmetry constraints, as explicitly shown in the Supplementary Information.\\
In pristine \Ca, the vertical mirror symmetry is globally preserved within the lattice, and the corresponding mirror-parity quantum number governs the selection rules for magnetic excitations and their coupling to external probes, including Raman scattering. In the full Hamiltonian $\mathcal{H}$, this symmetry is properly defined by the operator $\mathbb{M} \equiv M \otimes \prod_{j} \mathcal{M}_b^j$, which combines the local action $\mathcal{M}_b^j$ on site $j$ with the spatial mirror $M$ mapping the coordinates as $(x,y) \xrightarrow{M} (-y,-x)$. 
Consequently, two symmetry-distinct one-magnon branches, $\lvert \psi_-\rangle$ and $\lvert \psi_+ \rangle$, are expected; these branches are odd and even, respectively, under the global mirror operation, such that $\mathbb{M} \lvert \psi_- \rangle = -\lvert \psi_- \rangle$ and $\mathbb{M} \lvert \psi_+ \rangle =  \lvert \psi_+ \rangle$. An explicit demonstration of this symmetry based construction is provided in the Supplementary Information. Assuming the local terms dominate, we perform a perturbative expansion in the exchange coupling and derive the energies of the two magnons relative to the ground-state energy $E_0$, which at leading order in $\delta_b$ and $\delta_c$ are expressed as

\begin{equation}
E_- - E_0
= f_0 
+\frac{2\delta_b^2}{E_T}
+\left(2-\frac{E_T}{J_{xx}+J_{xy}}\right)
\frac{\delta_c^2\delta_b^2}{E_T^3},
\end{equation}

\begin{equation}
E_+ - E_0
= f_0
+\frac{4\delta_b^2}{E_T} 
+\left(4+\frac{2E_T}{J_{xx}+J_{xy}}\right)
\frac{\delta_c^2\delta_b^2}{E_T^3},
\end{equation}

with $f_0=E_T
-2\left(J_{xx}+J_{xy}\right)
-\delta_c^2/\left[4\left(J_{xx}+J_{xy}\right)\right]$. These expressions for the one-magnon modes naturally account for the energy shift induced by an increase in the anisotropy magnetic energy $E_T$, as well as for the emergence of split magnon modes, whose energy separation depends on both the magnitude of the octahedral distortions and the strength of the magnetic exchange.
%
%

To describe the Raman spectra, we adopt the LF framework~\cite{Fleury1968}, with the Raman scattering operator being defined as

\begin{equation}
R \propto \sum_{\langle i,j \rangle}
(\boldsymbol{\epsilon}_{\mathrm{in}}\!\cdot\!\mathbf{r}_{ij})
(\boldsymbol{\epsilon}_{\mathrm{out}}\!\cdot\!\mathbf{r}_{ij})\,
\mathcal{H}_{ij},
\label{eq:Raman_op}
\end{equation}
where $\boldsymbol{\epsilon}_{\mathrm{in}}$ and $\boldsymbol{\epsilon}_{\mathrm{out}}$ are the incoming and outgoing polarization vectors, and $\mathcal{H}_{ij}$ denotes the Hamiltonian on bond $\langle i,j\rangle$, 
which includes both exchange interactions $\mathcal{H}_x$ and $\mathcal{H}_y$~\cite{Fleury1968}. 

When the light polarizations are confined to the \textit{ab} plane, $\boldsymbol{\epsilon}_{\mathrm{in(out)}}$ can be specified by their in-plane angles with respect to a crystallographic axis, such as the $a$ axis. In the $B_{1g}$ geometry adopted in the experiments, the Raman operator reduces to
\begin{equation}
R_{B_{1g}} = 
\sum_{\langle i,j\rangle_x} \mathcal{H}_{ij}
- \sum_{\langle i,j\rangle_y} \mathcal{H}_{ij},
\label{eq:Raman_B1g}
\end{equation}

In pure \Ca, only a one-magnon branch is observed in Raman scattering within the $B_{1g}$ symmetry channel, as schematically shown in Fig.~\ref{fig:ConceptFig}b. 
This selective visibility is determined by two factors: 
(i) the symmetry of the Raman tensor, which transforms as odd under the mirror operation $\mathbb{M}$, and 
(ii) the symmetry of the magnetic exchange Hamiltonian, which, in contrast, is even under $\mathbb{M}$. As a result, only a single one-magnon branch, whose mirror parity is compatible with the Raman tensor (i.e., odd) can couple to light in the $B_{1g}$ channel. 
The other branch, which is even under $\mathbb{M}$, remains Raman inactive in this symmetry channel and allowed in the $A_{g}$ symmetry. However, the experimental observations here and in the literature \cite{souliou_raman_207} indicate that this magnetic mode is fully obscured in the $A_{g}$ channel due to the spin--phonon coupling. This is commonly observed in magnetic materials because fully symmetric magnetic excitations hybridize strongly with $A_{g}$ phonons, leading to pronounced screening and damping of the spectral-weight.
The appearance of an additional $B_{1g}$ Raman-active magnon in Mn-substituted samples thus signals a lowering of the local symmetry and the onset of mixed mirror-parity character in the magnetic excitations.

The substitution of Ru by Mn in \CaMn\  acts as a local symmetry-breaking perturbation. The observation of sharp, coherent Raman peaks throughout the sample indicates that this distortion does not arise from isolated small clusters, but rather acts as a global perturbation. This is further corroborated by Energy Dispersive Spectroscopy (EDS) mapping analysis (see Supplementary Information), which confirms the stoichiometric accuracy and a spatially homogeneous Mn distribution at the sub-micrometric level. We can therefore assume that Mn ions locally relax the octahedral compression around the impurity sites and thereby disrupt the global mirror symmetry $\mathbb{M}$ of the crystal lattice. This perturbation modifies both the crystal-field environment and the superexchange pathways, effectively mixing magnon states of different mirror parity. As a result of Mn substitution, the mirror operation $\mathbb{M}$ ceases to be an exact symmetry of the magnetic lattice. Consequently, the one-magnon excitations can no longer be classified by a well-defined mirror parity. The breaking of this symmetry lifts the selection rules that protect certain excitations, thereby activating mixed-parity one-magnon modes that are otherwise hidden in the high-symmetry phase (Fig.~\ref{fig:ConceptFig}c).

To substantiate these symmetry-based results 
we perform exact-diagonalization (ED) calculations of the Raman spectrum on finite clusters, with various distributions of  Mn impurities. The calculations are carried out using the full pseudospin Hamiltonian $\mathcal{H}=\mathcal{H}_{loc}+\mathcal{H}_x+\mathcal{H}_y$. The Raman operator $R_{B_{1g}}$ is implemented according to Eq.~\eqref{eq:Raman_B1g}, and the Raman intensity spectral function is given by 
\begin{equation}
I(\omega) = \sum_n |\langle n|R_{B_{1g}}|0\rangle|^2\,\delta(\omega - E_n + E_0) \,.
\end{equation}

In Fig. 4a-b, we present representative cases for both the undoped and single-impurity configurations, and the two-impurity configurations, respectively. Our focus is on the lowest-energy magnon mode, which is most relevant for the experimental observations. In the undoped case, a single one-magnon mode can be reproduced at approximately 12 meV assuming the following model parameters, $J_{xx}=5.2$ meV, $J_{xy}=3.1$ meV, $J_{z}=2$ meV, $E_T=30$ meV which are compatible with the magnetic properties of Ca$_2$RuO$_4$ \cite{GretarssonPRB2019,das_spin-orbital_2018}.

The introduction of an impurity breaks the vertical mirror symmetry by inducing a short-range anisotropic structural distortion around the impurity, consistent with the orthorhombicity of the crystal lattice. Depending on the magnitude of this local structural change, the strength of the mirror-symmetry breaking can be tuned, resulting in a shift of the emerging double-peak magnon structure to higher energies (Fig. 4a).
\\
Increasing the number of impurities, e.g. from one to two as in Fig. 4b, modulates the relative intensity of the two peaks and redistributes the magnon spectral weight, depending on the positions of the impurities relative to the magnetic sublattice and the gradient profile of the structural potential. When averaging over different impurity configurations, these effects produce a shift in the magnon energies and a finite broadening of the Raman spectral intensity. 
The resulting outcomes can reproduce the qualitative changes of the Raman spectra observed in the experiments as a function of the Mn concentration.


Finally, we discuss the implications of the angular polarization dependence of the Raman intensity for the mixed-parity magnons. As shown in Fig.~2, the one-magnon peaks exhibit a non-trivial angular dependence upon the polarization angle, providing crucial insight into the symmetry properties of the excitations. We first consider the pristine Ca$_2$RuO$_4$ system, where the $B_{1g}$ one-magnon mode deviates significantly from the standard four-fold $d$-wave-like intensity profile predicted by the non-resonant LF theory (see Figs.~2c and 2g). This deviation is particularly pronounced at an incident energy of 2.2 eV, suggesting the breakdown of the non-resonant approximation due to the activation of specific resonant scattering channels. Far from resonance, the Raman operator $\mathcal{R}$ is governed by the LF theory. As highlighted by Eq.~(5), this operator is symmetric under the exchange of incident and scattered polarization vectors. Near resonance, however, the perturbative expansion underlying the LF approximation breaks down. In this regime, an antisymmetric contribution to the Raman tensor---arising from intermediate-state transitions that break the exchange symmetry of the incoming and outgoing photon polarizations---emerges and acquires significant spectral weight~\cite{shastry1990}. We point out that including these resonant terms preserves the validity of our previous analysis: while the standard LF theory robustly explains the emergence of impurity-induced hidden magnons, adding these resonant effects is a necessary refinement to precisely reproduce the polarization-angle dependence of the intensity.
\\
Defining $\theta$ as the angle between the incident polarization vector and e.g. the $a$ crystallographic axis, the effective Raman operator in the $B_{1g}$ configuration is modeled as:
\begin{equation}
    \mathcal{R} \propto \frac{1}{2} \cos(2\theta) R_{LF} + R_{res},
    \label{eq:Raman_pure}
\end{equation}
where the first term represents the symmetric LF contribution, while $R_{res}$ denotes the $\theta$-independent resonant term. 
This resonant term, $R_{res}$, arises from transitions to intermediate states within the explicit Raman scattering amplitude~\cite{shastry1990,Devereaux2007}. Under resonant conditions, these intermediate transitions activate the antisymmetric components of the Raman tensor. Being rotationally invariant, this antisymmetric term does not modulate with $\theta$. Consequently, under resonant conditions, the total Raman intensity is described by a constant background superimposed on the standard symmetric modulation. This profile is shown in Fig.~\ref{fig:angular_model}a, which displays the calculated intensity obtained by setting the matrix elements of the operator in Eq.~(\ref{eq:Raman_pure}) to values that reproduce the experimental intensity ratio between the two polarization configurations rotated by 90$^{\circ}$, for the specific case at 2.2~eV (see Figs.~2c and 2g).\\
In the Mn-doped system, the scenario is fundamentally modified by the local breakdown of the lattice mirror symmetry, $\mathbb{M}$, induced by impurities. Experimentally, we observe two distinct one-magnon peaks with approximately two-fold symmetries that are mutually phase-shifted by $90^\circ$ (see Figs.~2d, 2e, 2h, 2i). This phenomenology indicates two concurrent effects driven by symmetry breaking: the hybridization between one-magnon states and resonance effects.\\
First, we note that the dopant-induced mirror symmetry breaking, hybridizes the original modes into mixed-parity states. Specifically, the two resulting one-magnon eigenstates, denoted as $|n_1\rangle$ and $|n_2\rangle$, arise from the parity mixing of the odd (allowed) $|\psi_-\rangle$ state and the even (hidden) $|\psi_+\rangle$ state, defined by a mixing angle $\alpha$:
\begin{equation}
    \begin{pmatrix} |n_1\rangle \\ |n_2\rangle \end{pmatrix} = 
    \begin{pmatrix} \cos\alpha & \sin\alpha \\ -\sin\alpha & \cos\alpha \end{pmatrix} 
    \begin{pmatrix} |\psi_-\rangle \\ |\psi_+\rangle \end{pmatrix}.
    \label{eq:mixing}
\end{equation}
The mirror-plane breaking must be imprinted on the spatial symmetry of the new eigenmodes, lowering their symmetry to a strongly anisotropic, twofold character. To directly probe this symmetry reduction, we employed polarization-resolved Raman spectroscopy. Because the Raman tensor reflects the underlying symmetry of the excitations, tracking the angular dependence of the polarization allows us to map this reconstruction. This provides a direct optical readout of the mixed-parity states, through their distinctive twofold orthogonal response, as shown in Figs.~2d, 2e, 2h, and 2i for moderate Mn concentrations.\\
Second, and crucial for the detection of these mixed states, the breakdown of mirror symmetry modifies the Raman operator itself. While in the pristine case the operator $\mathcal{R}$ is strictly odd with respect to $\mathbb{M}$, allowing transitions only to the odd one-magnon state $|\psi_-\rangle$ (since the ground state is even), the presence of impurities relaxes this constraint. The impurity-induced operator, $\mathcal{R}_{imp}$, is no longer constrained to be parity-definite. It acquires contributions, denoted as $R^+$ and $R^+_{res}$, that are even with respect to $\mathbb{M}$, a channel formally forbidden in the pure system. These new terms enable a direct coupling to the even component $|\psi_+\rangle$ of the mixed eigenstates. The generalized impurity-modified operator can thus be written as:
\begin{equation}
    \mathcal{R}_{imp} \propto \frac{1}{2}\cos(2\theta)(R^- + R^+) + (R^-_{res} + R^+_{res}),
    \label{eq:Raman_imp}
\end{equation}
where $R^{-}$($R^{+}$) and $R^{-}_{res}$($R^{+}_{res}$) yield non-vanishing amplitudes on the $|\psi_-\rangle$ ($|\psi_+\rangle$) state only. The theoretical angular dependence derived from Eq.~\eqref{eq:Raman_imp} is shown in Fig.~\ref{fig:angular_model}b. This profile was obtained by using matrix elements for the Raman operator that are consistent with the experimental intensity ratios of the two one-magnon peaks for the $n$-doped system, specifically for the representative case of $x = 5\%$ at 2.2~eV ( Figs.~2e and 2i). Furthermore, a mixing angle $\alpha = \pi/8$ was chosen for the mixed-parity state, as it provides a good agreement with the experimental data. The model successfully reproduces the observed 90$^\circ$ phase shift between the intensities of the two modes. Therefore, the observed orthogonality of the magnon intensity profiles is a signature of mixed-parity magnons; it is enabled by the impurity-induced relaxation of selection rules due to mirror symmetry breaking, with the angular modulation being further influenced by resonant scattering effects.


\section{Discussion and conclusions}
We have shown that partial substitution of Ru with Mn in $\mathrm{Ca_2RuO_4}$ reconstructs the magnon spectrum and activates one-magnon excitations that are hidden in the undoped material. This effect arises from local mirror-symmetry breaking of the spin-orbital configuration, induced by structural distortions of the $\mathrm{RuO_6}$ octahedra near the dopant, and demonstrates the critical role of spin-orbit-lattice entanglement in shaping collective magnetic excitations.
Our findings highlight that spin-orbital magnons are highly sensitive to local structural and electronic perturbations, providing a pathway to control magnetic dynamics beyond conventional spin-only physics. 
More broadly, the mechanism uncovered here suggests that hidden magnons in a wide range of spin-orbit correlated materials, especially those with large local magnetic moments, could be selectively accessed through dopants, strain, or other pathways for inducing lattice distortions, such as applied electric fields, thereby opening new opportunities to engineer magnonic spectra for quantum and spintronic devices.
Finally, our results demonstrate that the mixed--parity character of the coupled magnons gives rise to a re\-mar\-ka\-ble polarization dependence of the magnetic excitations. This manifests as a changeover from fourfold to twofold rotational symmetry, producing a pronounced nematic-like polarization response. The observed symmetry reduction also reflects the intricate interplay between resonant and nonresonant scattering channels and highlights how spin--orbit--lattice entanglement can be exploited to generate magneto-optical effects based on both the symmetry and polarization characteristics of collective magnetic modes in correlated materials.


\section{acknowledgments}
M.C., F.F., R.F., A.G., M.L. and A.V. acknowledge partial support by the Italian Ministry of Foreign Affairs and International Cooperation, grant KR23GR06. F.F. and F.G. acknowledge support from the Italian Ministry of University and Research (MUR) under the National Recovery and Resilience Plan (NRRP), Call PRIN 2022, funded by the European Union - NextGenerationEU, Mission 4, Component 2, Grant No. 2022TWZ9NR (STIMO)-CUP B53D23004560006.
M.C. and A.V. acknowledge support by Italian MUR PRIN 2022 under the Grant No. 2022LP5K7 (BEAT). M.C. and F.F. acknowledge support from
PNRR MUR project PE0000023-NQSTI.
W.B. acknowledges support by Narodowe Centrum Nauki (NCN, National Science Centre, Poland) Project No. 2021/43/B/ST3/02166 and by the Foundation for Polish Science project MagTop no. FENG.02.01IP.05-0028/23 co-financed by the European Union from the funds of Priority 2 of the European Funds for a Smart Economy Program 2021--2027 (FENG). D.W. acknowledges support from the Institute of Applied Physics of Seoul National University, from the ITRC Program through the IITP, and from the Global Research Development Center (GRDC) Cooperative Hub Program through the NRF, funded by the (MSIT) (Grant Nos. RS-2024-00437191 and RS-2023-00258359).

\section{Author Contributions}
D.W. and F.G. contributed equally to this manuscript.
M.C., R.F., F.F., C.K., and D.W. conceived the research project.
M.C., F.F., C.K., and D.W. designed the study, with contributions from R.F. and A.V.
R.F., A.G., M.L. and A.V. grew and characterized the Ca$_2$(Ru,Mn)O$_4$ single crystals.
D.W. carried out the Raman measurements and performed the data interpretation, F.F., F.G. and M.C. contributed to the data discussion. 
M.C., F.F., and F.G. proposed the mechanism of symmetry-breaking-induced magnons.
M.C., F.F., and F.G. developed the theoretical model, with contributions from W.B. 
W.B. carried out the Raman spectrum simulations, with contributions from M.C., F.F., and F.G.\\
M.C., F.F., F.G., and D.W. wrote the manuscript with contributions from all authors.
\\





\bibliography{biblio.bib}
\newpage

\begin{figure*}
\label{RamanPolar}
\centering
\includegraphics[width=16cm]{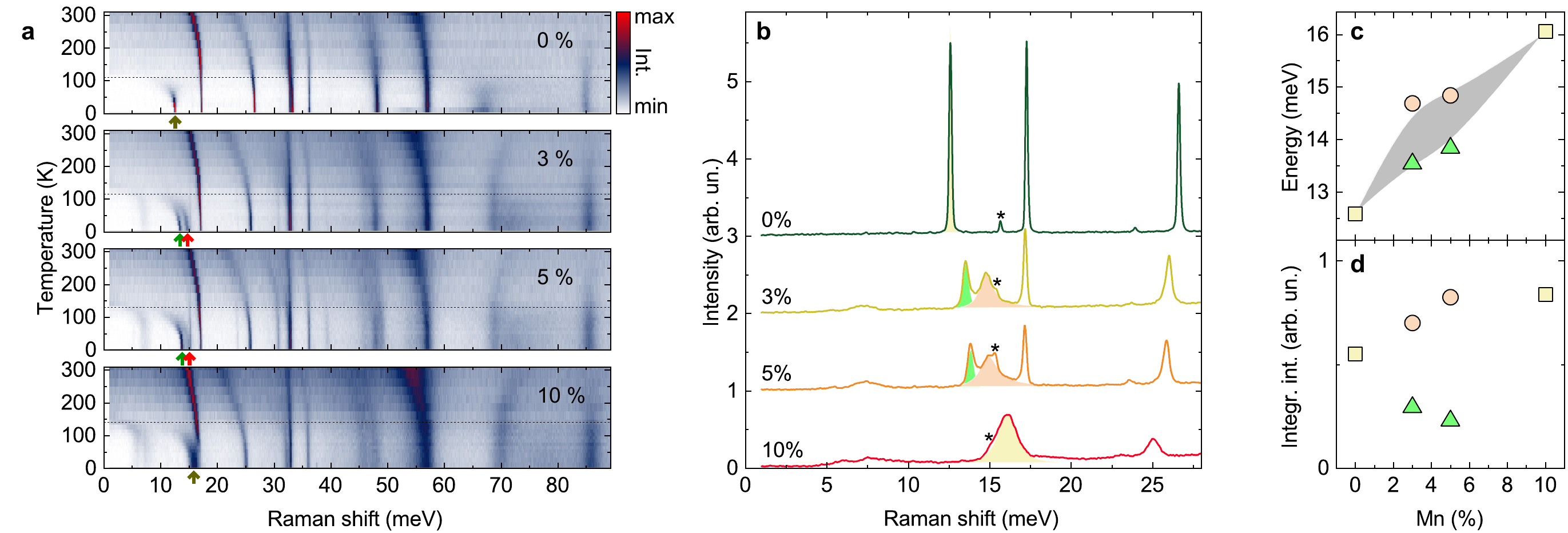}
\captionsetup[figure]{labelfont=bf, labelsep=period, name=Fig.}
\caption{\textbf{Temperature and doping dependence of magnons in Ca$_2$(Ru,Mn)O$_4$.}\\
\textbf{a}, Temperature-dependent Raman spectra of Ca$_2$(Ru,Mn)O$_4$ with 0\%, 3\%, 5\%, and 10\% Mn content measured in the $B_{1g}$ symmetry channel. Dashed lines mark the N\'{e}el temperatures. The arrows indicate one-magnon excitations. \textbf{b}, Individual Raman spectra measured at $T = 4$ K. One-magnon contributions are shaded. The black asterisks mark a phonon that partially overlaps with magnons. \textbf{c}, Magnon energies and \textbf{d}, their integrated intensities as a function of Mn content.}
\end{figure*}

\begin{figure*}
\label{RamanPolar}
\centering
\includegraphics[width=16cm]{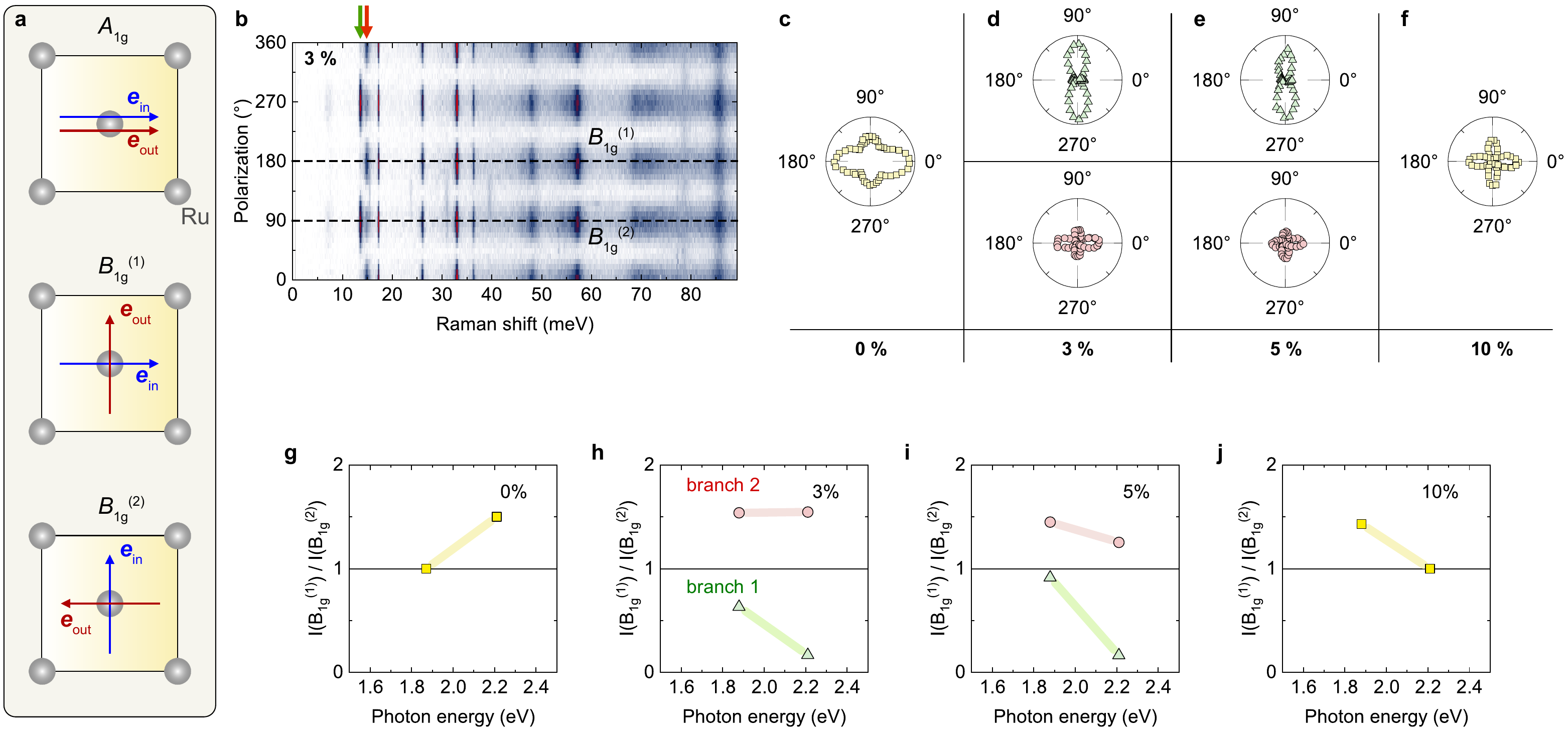}
\captionsetup[figure]{labelfont=bf, labelsep=period, name=Fig.}
\caption{\textbf{Evolution of the polarization-resolved magnon spectrum in Ca$_2$(Ru,Mn)O$_4$}. \textbf{a}, Schematic sketch of Raman scattering configurations in different symmetry channels. The polarization vectors of incident and outgoing light (blue and red, respectively) are indicated with respect to the Ru atoms (grey spheres). \textbf{b}, Polarization-resolved Raman spectra of Ca$_2$(Ru,Mn)O$_4$ (Mn = 3\%) measured at $T = 4$ K with $\boldsymbol{\epsilon}_{\mathrm{in}} \perp \boldsymbol{\epsilon}_{\mathrm{out}}$. The two individually resolved one-magnon excitations are marked by a green (lower branch) and a red (higher branch) arrow. \textbf{c-f}, Extracted linecuts of the one-magnon mode intensity as a function of light polarization. In panels \textbf{c} and \textbf{f} the yellow squares denote the single peaks observed in samples with 0\% and 10\% Mn content. In panels \textbf{d} and \textbf{e} the green triangles trace the lower branches, while the red circles mark the higher branches in samples with 3\% and 5\% Mn content. \textbf{g-j}, Degree of asymmetry, quantified as the intensity ratio of lobes, as a function of photon energy for all four samples.}
\end{figure*}

\begin{figure}[ht!]
\includegraphics[width=7.74cm]{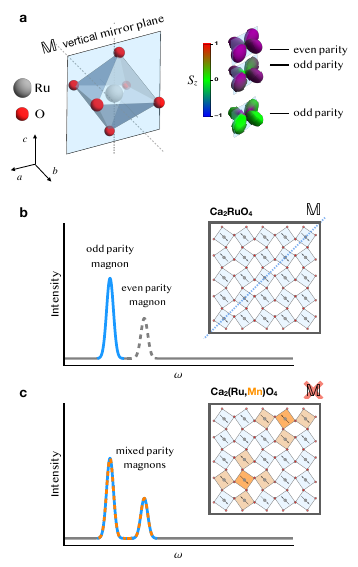}
\captionsetup[figure]{labelfont=bf, labelsep=period, name=Fig.}
    \caption{\textbf{Spin--orbital states and mirror--symmetry breaking in Ca$_2$RuO$_4$.}\\
    \textbf{a}, Schematic of the of the three lowest-lying local spin--orbital states stabilized by SOC and octahedral distortions in \Ca. The {RuO$_6$} octahedra tilting relative to the $b$ axis mixes orbital characters, splitting the multiplets into states of definite parity with respect to the diagonal vertical mirror ($ac$ plane perpendicular to the $b$ axis). The energy state has odd parity, while the higher-energy doublet comprises even and odd parity states. The panel illustrates a representative case associated with a specific sign of the parameter $\delta_c$ as related to the staggering of the octahedral rotations about to the $c$-axis; for the opposite sign, the parity of these two states is inverted. \textbf{b},\textbf{c}, Schematic of the evolution of the magnon spectrum for the antiferromagnetic state from the undoped (panel \textbf{b}) to the Mn-doped configuration (panel \textbf{c}), where Mn substitution breaks the diagonal vertical mirror ($\mathbb{M}$). This symmetry breaking activates a one-magnon mode in the cross polarization channel that is silent in the Ca$_2$RuO$_4$ compound. Here, we focus on the excitations of the Ru magnetic moments, with the oxygen degrees of freedom projected out. This allows applying a mirror transformation to the Ru lattice and the associated spin configuration, as illustrated in \textbf{b}.}
    \label{fig:ConceptFig}
\end{figure}

\begin{figure*}
  \centering
  \includegraphics[width=1.0\textwidth]{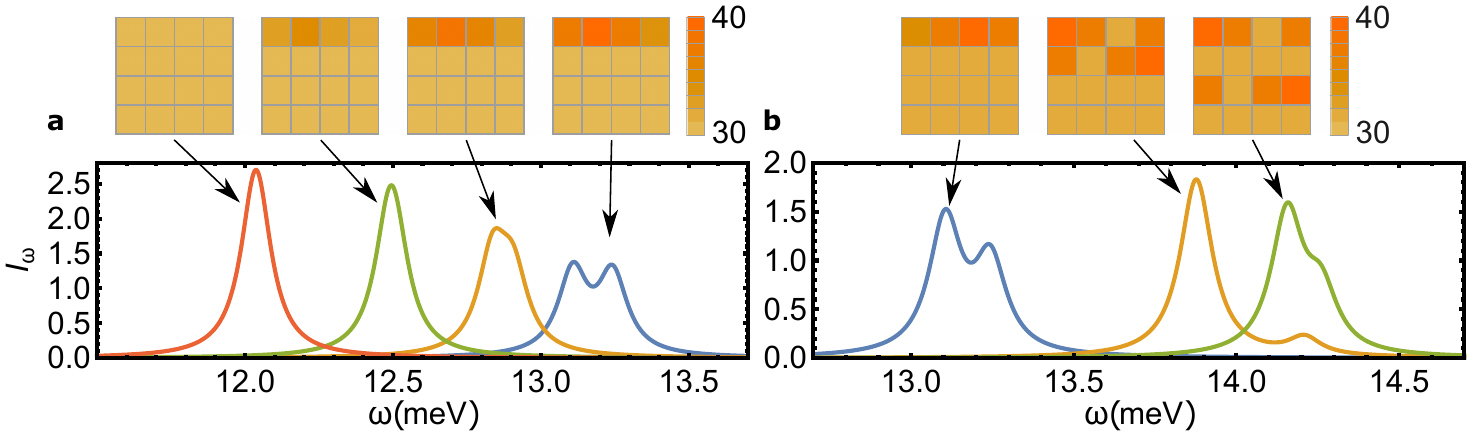} 
  \captionsetup[figure]{labelfont=bf, labelsep=period, name=Fig.}
  \caption{\textbf{Raman spectral function of a cluster with anisotropic magnetic perturbations}.
  Raman spectral function evaluated for a $4\times4$ cluster with periodic boundary conditions and in the presence of inhomogeneous anisotropic magnetic energy $E_T$ nearby the substituted site (corresponding to the darkest orange square in the schematic cluster). \textbf{a}, Undoped and single-impurity configurations; \textbf{b}, two-impurity configurations. The parameters of the model are $J_{xx}=5.2$, $J_{xy}=3.1$, $J_{z}=2$, $E_T=30$ (in units of meV) at host sites. At the impurity sites the anisotropic magnetic energy $E_T$ is modified to acquire the value $E_T=40$ and is also varied at the neighboring sites of the impurity, as illustrated by the contour map.
  For clarity we employ Lorentzian distributions for the spectral function with broadening factor $\eta=0.06$ meV.
  }
\end{figure*}

\begin{figure}[t]
    \centering
 \includegraphics[width=12cm]{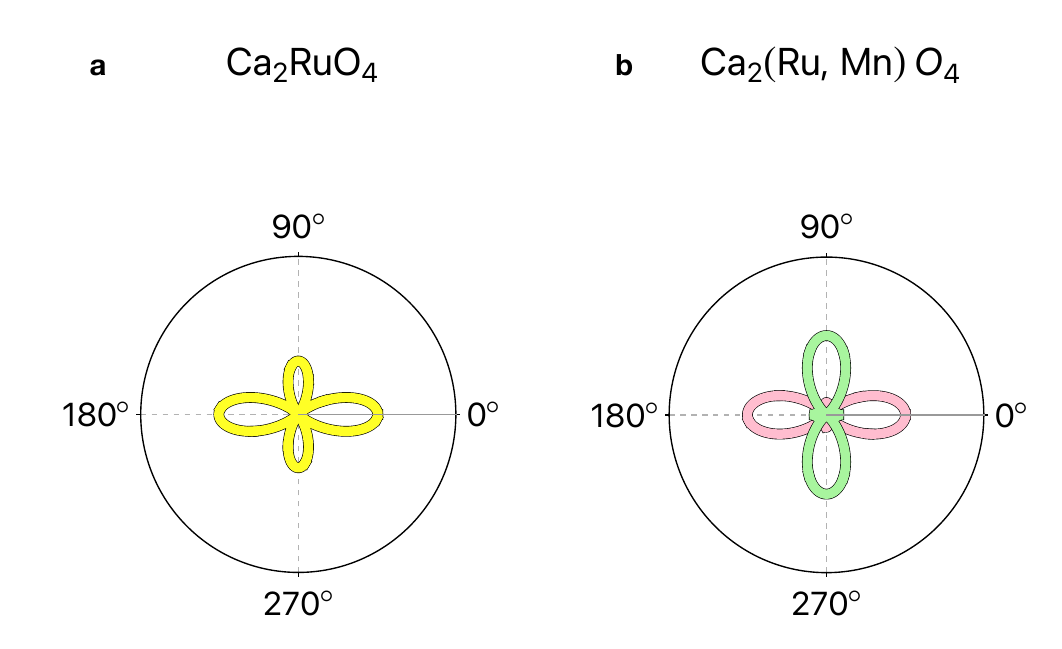} 
  \captionsetup[figure]{labelfont=bf, labelsep=period, name=Fig.}
    \caption{
    \textbf{Calculated angular Raman intensity profiles highlighting symmetry breaking in Mn-doped Ca$_2$RuO$_4$.}
    Computed angular dependence of Raman Intensity. 
    \textbf{a}, Pristine Ca$_2$RuO$_4$ system: The intensity profile (yellow line) follows the superposition of a symmetric $B_{1g}$ term ($\cos 2\theta$) and an isotropic resonant background, resulting in an anisotropic 4-fold symmetry. The calculated profile assumes a representative ratio of 1.5 between the 90$^{\circ}$ phase-shifted intensities, corresponding approximately to the maximum lobe intensity ratio shown in Fig.~2g. Further details on the specific values of the matrix elements derived from Eqs.~(\ref{eq:mixing})-(\ref{eq:Raman_imp}) can be found in the Supplementary Information.
    \textbf{b}, Mn-doped system: The symmetry breaking leads to two distinct modes, $|n_1\rangle$ (green) and $|n_2\rangle$ (pink). The interplay between parity mixing and resonant scattering results in two-fold symmetric lobes that are phase-shifted by $90^\circ$. The angular profile was calculated using a mixing angle $\alpha = \pi/8$ and matrix elements optimized to reproduce the $x=5\%$ ($2.2$~eV) intensity ratios from Fig.~2i. Further details are provided in the Supplementary Information.}
    \label{fig:angular_model}
\end{figure}

\end{document}